\documentclass{article}

\usepackage{arxiv}

\usepackage[utf8]{inputenc} 
\usepackage[T1]{fontenc}    
\usepackage{url}            
\usepackage{booktabs}       
\usepackage{amsfonts}       
\usepackage{nicefrac}       
\usepackage{microtype}      
\usepackage{graphicx}
\usepackage{doi}
\usepackage[all]{xy}
\usepackage{setspace}
\usepackage{xcolor}
\usepackage{tikz}
\usetikzlibrary{cd}

\title{Classical and Multiscale Non-equilibrium Thermodynamics}

\author{Miroslav Grmela\\
\'{E}cole Polytechnique de Montr\'{e}al,
  C.P.6079 suc. Centre-ville,\\
 Montr\'{e}al, H3C 3A7,  Qu\'{e}bec, Canada\\
\And Michal Pavelka \\
 Mathematical Institute, Faculty of Mathematics and Physics, Charles University,\\ 
 Sokolovsk\'{a} 83, 18675 Prague, Czech Republic\\
 Corresponding author: pavelka@karlin.mff.cuni.cz
}

\newcommand{\shorttitle}{Classical and Multiscale Non-equilibrium Thermodynamics}

\counterwithin*{equation}{section} 
\usepackage{amsmath,amssymb,amsfonts,latexsym,float,graphics,epsfig}

\newcommand{\RR}{{\boldmath \mbox{$R$}}}

\newcommand{\JJ}{{\boldmath \mbox{$J$}}}

\newcommand{\uu}{{\boldmath \mbox{$u$}}}
\newcommand{\bb}{{\boldmath \mbox{$b$}}}

\newcommand{\rr}{{\boldmath \mbox{$r$}}}

\newcommand{\cc}{{\boldmath \mbox{$c$}}}

\newcommand{\vv}{{\boldmath \mbox{$v$}}}

\begin{document}
\maketitle

\begin{abstract}
Classical and multiscale non-equilibrium thermodynamics have  different histories  and  different objectives. In this Note we explain the differences and review some topics in which the multiscale viewpoint of mesoscopic time evolution of macroscopic systems helped to advance the classical non-equilibrium thermodynamics. Eventually, we illustrate the Braun-Le Ch{\^a}telier principle in dissipative thermodynamics.
\end{abstract}

\tableofcontents

\doublespace

\section{Introduction}

The classical non-equilibrium thermodynamics grew out of  combining  the classical fluid mechanics with the assumption of local  equilibrium and  the requirement of satisfying  the second law of thermodynamics \cite{dgm,prigogine-tip}. The multiscale non-equilibrium thermodynamics grew  out of an attempt to extract from the Boltzmann kinetic equation a mathematical structure that expresses both the underlying mechanics of microscopic particles composing macroscopic systems and the second law of thermodynamics. Governing equations of the classical non-equilibrium thermodynamics as well as governing equations of other mesoscopic theories of macroscopic systems are then particular realizations of
the resulting "abstract Boltzmann equation" (called GENERIC - an acronym for General Equation for Non-Equilibrium Reversible-Irreversible Coupling) \cite{go,og,hco,pkg}.

In Section \ref{sec2}, we recall the classical non-equilibrium thermodynamics, in Section \ref{sec3} we recall GENERIC, and in Section \ref{sec4} we review some topics in which a combination of both viewpoints of mesoscopic dynamics advanced the classical non-equilibrium thermodynamics. The topics are: (i) Time reversible part of the time evolution in the classical and extended  non-equilibrium thermodynamics, Hamiltonian Grad hierarchy.  (ii) Local equilibrium assumption, the second law  in extended  non-equilibrium thermodynamics.  (iii) Nonlinear dissipation. (iv) Role of the Braun-Le Chatelier in  the time evolution that takes place in the space of constitutive relations.

\section{Classical non-equilibrium thermodynamics}\label{sec2}

\subsection{Balance laws and the second law of thermodynamics}
Euler transposed  Newton mechanics of particles to continuum  \cite{euler}. The resulting Euler equation supplemented with two equations expressing the conservations of mass and energy  and with friction are  governing equations of the classical Navier-Stokes fluid mechanics. The state variables are the hydrodynamic fields $x^{(hyd)}=(\uu(\rr),\rho(\rr),e(\rr))$ denoting the momentum, mass, and energy respectively, $\rr\in\mathbb{R}^3$ is the position vector.

This theory is not, however, complete since the systems investigated in fluid mechanics are not only particles obeying Newton's mechanics. They are macroscopic systems that, if left undisturbed, reach an equilibrium state at which their behavior is well described by the classical equilibrium thermodynamics \cite{dgm}. 

Continuum mechanics has to display this observation. This is done by extending the interpretation of the energy $e(\rr)$ and assuming the local thermodynamic equilibrium. By the former modification we mean that the total energy $e(\rr)$ is the sum of the mechanical energy (that is the kinetic energy of the continuum $\frac{\uu^2(\rr)}{2\rho(\rr)}$) and an internal energy $\epsilon(\rr)$ that is the energy of the internal structure of the continuum (that is left unspecified). The latter modification introduces a new field $s(\rr)$ denoting the local entropy. This new field is not an additional state variable, it is not an independent field, it is a function of the fields $(\rho(\rr), \epsilon(\rr))$. Moreover, the function is known if we know the equilibrium fundamental thermodynamic relation of the fluid under investigation in the equilibrium sate reached by following solutions of the governing equations \cite{callen}. 

The field $s(\rr)$ depends on $(\rho(\rr), \epsilon(\rr))$ for all position vectors $\rr$ in the same way as the equilibrium entropy per unit volume depends on the equilibrium mass per unit volume and the equilibrium energy per unit volume in the equilibrium state. With such local equilibrium assumption, it is now necessary to guarantee that the local equilibrium approaches (by following solutions of the time evolution equations of the continuum mechanics in the absence of external forces) the total thermodynamic equilibrium with the same relation between entropy, mass, and energy (all per unit volume). This requirement is formulated as the second law of thermodynamics
\begin{subequations}
\begin{eqnarray}
\frac{\partial s(\rr)}{\partial t}&=&-\partial_iJ^{(s)}_i(\rr)+\sigma(\rr)\nonumber \\
\sigma(\rr)&\geq& 0 \label{2law}
\end{eqnarray}
\end{subequations}
which gives the total entropy multiplied by $-1$, $S=-\int d\rr s(\rr)$,  the role of the Lyapunov function for the approach towards equilibrium. We use the following notation in Eq.(\ref{2law}): $\JJ^{(s)}(\rr)$ is the entropy flux, $\sigma(\rr)$ is the entropy production, $\partial_i=\frac{\partial}{\partial r_i}$, where $\rr=(r_1,r_2,r_3)$, and the summation convention over repeated indices is used.

The governing equations of the classical non-equilibrium thermodynamics  are:  Eq.(\ref{2law}) and
\begin{subequations}\label{eqfm}
\begin{eqnarray}
\frac{\partial u_i}{\partial t}&=& -\partial_jJ^{(u)}_{ij}\\
\frac{\partial \rho}{\partial t}&=& -\partial_iJ^{(\rho)}_{i}\\
\frac{\partial e}{\partial t}&=& -\partial_iJ^{(e)}_{i}
\end{eqnarray}
\end{subequations}
where $\JJ^{(u)}, J^{(\rho)}, J^{(e)}$ are fluxes of momentum, mass, and energy respectively.

These equations can become equations governing the time evolution of the state variables $x^{(hyd)}=(\uu(\rr),\rho(\rr),e(\rr))$ only if
\begin{equation}\label{cr}
s(x^{(hyd)}), \JJ^{(u)}(x^{(hyd)}), J^{(\rho)}(x^{(hyd)}), J^{(e)}(x^{(hyd)}), J^{(s)}(x^{(hyd)}), \sigma(x^{(hyd)})
\end{equation}
are specified. The quantities (\ref{cr}) specified as functions of $x^{(hyd)}=(\uu(\rr),\rho(\rr),e(\rr))$  are called \textit{constitutive relations}. The individual nature of the macroscopic systems under investigation is expressed in the classical fluid mechanics in the constitutive relations.

\subsection{Towards complex fluids}
The setting of the classical non-equilibrium thermodynamics  sketched above has been found insufficient for   complex fluids (as for example polymeric fluids and suspensions) and for fluids observed on small scales (for instance heat transfer in electronic devices) \cite{jou-eit,david-liliana}. Complex internal structure of the  fluids evolves in time on the same timescale as the hydrodynamic fields. Consequently, the setting (\ref{eqfm}), (\ref{2law}) has to be extended by adopting extra state variables (extra fields, we denote them by the symbol $\xi(\rr)$) into the  hydrodynamic fields $x^{(hyd)}$. The extra fields  characterize states of the internal structure.

Mathematical formulation of the  extended continuum mechanics thus consists of (\ref{eqfm}),(\ref{2law}),
\begin{equation}\label{extra}
\frac{\partial \xi}{\partial t}=\mathcal{R}(x^{(exthyd)})
\end{equation}
and constitutive relations for
\begin{eqnarray}\label{crex}
&&\xi(\rr), s(x^{(exthyd)}), \JJ^{(u)}(x^{(exthyd)}), J^{(\rho)}(x^{(exthyd)}),\nonumber \\ && J^{(e)}(x^{(exthyd)}), J^{(s)}(x^{(exthyd)}), \sigma(x^{(exthyd)}), \mathcal{R}(x^{(exthyd)})
\end{eqnarray}
where $x^{(exthyd)}=(\uu(\rr),\rho(\rr),e(\rr),\xi(\rr))$.

\section{Multiscale non-equilibrium thermodynamics}\label{sec3}

Multiscale non-equilibrium thermodynamics grew out of Boltzmann's investigation of the time evolution of the ideal gas \cite{boltzmann} and Arnold's investigation \cite{arnold} of the Hamiltonian structure on the Euler equation in fluid mechanics. The Boltzmann's investigation is historically the first attempt to formulate mathematically the time evolution of a macroscopic system displaying the experimentally observed approach to thermodynamic equilibrium at which the classical equilibrium thermodynamic holds. The Arnold investigation follows the Clebsch reformulation \cite{clebsch} of the Euler equation (Newton equation for the continuum) into the Hamiltonian form. In Arnold's formulation, the problem of casting the Euler equation into the Hamiltonian form becomes closely related to older results about Lie algebras and related to it non-canonical Hamiltonian structures \cite{poisson}. 

With this connection established, it was then possible to cast also other time reversible and non-dissipative parts of other mesoscopic time evolutions (in particular then the time evolutions arising in kinetic theories \cite{morrison-greene1980}) into noncanonical Hamiltonian form \cite{dv}. In the Boltzmann kinetic equation the part expressing the free flow is the Hamiltonian term in the time evolution. The second term on the right-hand side of the Boltzmann equation that expresses  the contribution of  collisions has been cast \cite{grcontmath} into a form that is comparable in its abstractness to the Hamiltonian formulation of the free flow term by using Waldmann's view \cite{waldmann} of collisions as chemical reactions (as generalized gradient dynamics). The resulting reformulation of the Boltzmann equation in the form \cite{grcontmath}, 
\begin{equation}\label{generic}
\dot{x}=L E_x+\left[\Xi_{x^*}(x,x^*)\right]_{x^*=S_x}
\end{equation}
has been called GENERIC (an acronym for General Equation for Non Equilibrium Reversible-Irreversible Coupling). Non-canonical Hamiltonian dynamics was also been combined with  gradient dynamics  in \cite{dv},\cite{mor}, \cite{kauf}.

The following notation is used in (\ref{generic}). The symbol $x$ denotes state variables. In the kinetic theory $x$ is the one particle distribution function $f(\rr,\vv)$, where $\rr,\vv$ is the position vector and momentum of one particle respectively. In the classical fluid mechanics $x=x^{(hyd)}=(\uu(\rr),\rho(\rr),e(\rr))$, in kinetic theory $x=f(\rr,\vv)$, is the one particle distribution function, $(\rr,\vv)$ denotes the position vector and momentum of one particle. By the symbol $L$ we denote Poisson bivector expressing Hamiltonian kinematics of $x$. A convenient way to present $L$  is to present it in the form of the Poisson bracket $\{A,B\}=\langle A_x,LB_x\rangle$, where $A$ and $B$ are real valued and sufficiently regular functions of $x$, where subscripts stand for functional derivatives, $A_x=\frac{\delta A}{\delta x}$, and similarly in $B_x$. 
In the case when $x$ is an element of an infinite dimensional space (as it is the case of kinetic theory and fluid mechanics), the derivative is an appropriate functional derivative, which turns to the usual partial derivative in finite dimensions. A mathematical definition of the Poisson bracket can be find for instance in \cite{pkg}. 

Here we list properties of Poisson brackets on an example the Poisson bracket
\begin{equation}\label{PB}
\{A,B\}=\int d\rr\int d\vv f(\rr,\vv)\left(\frac{\partial A}{\partial r_i}\frac{\partial B}{\partial v_i}-\frac{\partial B}{\partial r_i}\frac{\partial A}{\partial v_i}\right)
\end{equation}
expressing kinematics of the one particle distribution function $f(\rr,\vv)$, $\rr$ denoting the position and $\vv$ the momentum. We see from (\ref{PB}) that $\{A,B\}=-\{B,A\}$ (which means that $L$ is skew symmetric). We can also easily verify: (i)  that the bracket (\ref{PB}) is degenerate in the sense that $\{A,C\}=0$ for all $A$ is $C=\int d\rr\int d\vv \eta(f(\rr,\vv))$, where $\eta: \mathbb{R}\rightarrow\mathbb{R}$, (ii) that the Jacobi identity $\{\{A,B\},C\}+\{\{C,A\},B\}+\{\{B,C\},A\}=0$ holds. The function $C(f)$ satisfying $\{A,C\}=0 \forall A$ is called a Casimir. The function $\eta(f)$ remains in our discussions of the non-dissipative part of the time evolution unspecified. In the Boltzmann investigation the inclusion of the influence of collisions on the time evolution (i.e.inclusion of the consideration of the dissipative part of the time evolution) leads to the choice $\eta(f) = -k_B f(\ln(h^3 f)-1)$, see \cite{pkg}.

From the physical point of view, the bracket (\ref{PB}) is the mathematical formulation of the kinematics of one particle (expressed in the canonical Poisson bracket $\{a,b\}=\frac{\partial a}{\partial r_i}\frac{\partial b}{\partial v_i}-
\frac{\partial b}{\partial r_i}\frac{\partial a}{\partial v_i}$, where $a$ and $b$ are real valued functions of $(\rr,\vv)$) transposed from the space with elements $(\rr,\vv)$ to the space with elements $f(\rr,\vv)$. The transposition is made through the relation between the group of canonical transformations of $(\rr,\vv)$ and the Poisson bracket on the dual of the  Lie algebra corresponding to the group (see e.g. \cite{pkg}). The relation between the Poisson bracket (\ref{PB}) and the canonical Poisson bracket for one particle can also be seen by noting the (\ref{PB}) transforms into the canonical one-particle Poisson bracket if $f$ is the Klimontovich distribution function (i.e. a delta function) and $A(f)= \int d\rr\int d\vv a(\rr,\vv)f(\rr,\vv),  B(f)= \int d\rr\int d\vv b(\rr,\vv)f(\rr,\vv)$. Roughly speaking, the bracket (\ref{PB}) is an average of the canonical one-particle bracket.

The symbol $E(f)$ in (\ref{generic}) denotes the energy. In the Boltzmann analysis $E(f)=\int d\rr\int d\vv \frac{\vv^2}{2m}f(\rr,\vv)$ that is the average kinetic energy of one particle, $m$ is the mass of one particle. The symbol $S(f)$ denotes the entropy. In the Boltzmann analysis $S(f)=\int d\rr\int d\vv \eta(f)$. The symbol $\Xi$ denotes dissipation potential. This potential is required to satisfy the following properties: (i) $\Xi(x,0)=0 \forall x$, (ii) $\Xi(x,x^*)$ reaches its minimum at $x^*=0$ for all $x$, (iii) $\Xi(x,x^*)$ is in a small neighborhood of $x^*=0$ a convex function of $x^*$ for all $x$ (which means that in a small neighborhood of $x^*=0$ the dissipation potential looks like a quadratic functional $\Xi(x,x^*)= \langle x^*,\Lambda(x) x^*\rangle$, where $\Lambda$ is a positive definite linear operator). In the Boltzmann analysis $\Xi= \int d\rr\int d\vv\int d\vv'\int d\vv_1\int d\vv'_1 W \cosh X$, where $X=(f^*(\rr,\vv'_1)+f^*(\rr,\vv')-f^*(\rr,\vv_1)-f^*(\rr,\vv))/2k_B$, and $(\vv,\vv_1)\leftrightarrows (\vv',\vv'_1)$ is a collision transformation which is a  one-to-one transformation leaving unchanged the momentum $\vv+\vv_1$ and the energy $\vv^2+(\vv')^2$.

The GENERIC equation (\ref{generic}) becomes an equation governing the time evolution of a specific macroscopic system if
\begin{equation}\label{cr1}
x,L(x),\Xi(x,x^*),S(x),E(x),N(x)
\end{equation}
are specified. We note a resemblance  between the abstract setting of the classical non-equilibrium thermodynamics (\ref{2law}), (\ref{eqfm}) with the constitutive relations (\ref{cr}) and the abstract setting of mesoscopic dynamics (\ref{generic}) with (\ref{cr1}). In the context of the GENERIC equation (\ref{generic}) we do not call specifications of (\ref{cr1})  constitutive relations but a particular realization of GENERIC. This terminology is borrowed from mathematics (for example "a particular representation or realization of a group").

The main difference between the classical non-equilibrium thermodynamics recalled in the previous section and the multiscale non-equilibrium thermodynamics represented by (\ref{generic}) is the multiscale nature of the latter. The GENERIC equation (\ref{generic})  is an abstract framework for the time evolution of macroscopic systems on  mesoscopic levels of description. The state variables do not have to be hydrodynamic or generalized hydrodynamic fields, the multiscale thermodynamics is not fluid mechanics or extended fluid mechanics.

One of the advantages of the abstract formulation  is the possibility to use  insights and  results collected in the abstract setting and  in settings of  its particular realizations representing
well established mesoscopic theories in other mesoscopic theories. In the next section we present a few contributions to the classical and extended fluid mechanics from  considering them as particular realizations of GENERIC (\ref{generic}).

\section{Multiscale view of the classical non-equilibrium thermodynamics}\label{sec4}

In this section we recall a few issues in the classical and extended non-equilibrium thermodynamics and discuss them through eyes of the multiscale non-equilibrium thermodynamics.

\subsection{Non-dissipative  part of the time evolution}

The  local conservation laws (\ref{eqfm}), the local equilibrium assumption, and the second  law (\ref{2law}) provide a good guide to specify the constitutive relations (\ref{cr}) (i.e. to make an association between fluids under investigation and constitutive relations). More difficult is to specify constitutive relations (\ref{cr1}) in the extended non-equilibrium thermodynamics. The second law (\ref{2law}), in particular when it is complemented with ingenious procedures that show how to use it (e.g. \cite{liu1972}),  still provides a good guide for specifying the dissipative part of the time evolution in extended theories. The  setting of  local conservation laws (\ref{eqfm}) that guides the specification of the  non-dissipative part of the time evolution in extended theories ceases  however to be applicable since $\int d\rr \xi(\rr)$ is not  typically  conserved. The multiscale viewpoint remains  applicable and  provides   assistance.

The first and also the most difficult step in extensions of the classical non-equilibrium thermodynamics is the choice of  extra fields. From the physical point of view, the question is of what are the quantities that  most faithfully represent the aspects of the macroscopic system under consideration that cannot be ignored in its time evolution observed on the scale of continuum mechanics.  A strategy  to answer this question (from both the physical and the  mathematical point of view)  is to focus first on the "rules" of motion (i.e. on its group structure). For instance the motion of the classical continuum is a Lie group of transformations $\mathbb{R}^3\rightarrow \mathbb{R}^3$. The state variable (i.e. the momentum field $\uu(\rr)$) as well as the kinematics expressed in the Poisson bracket
\begin{equation}\label{PBu}
\{A,B\}=\int d\rr u_i\left(\frac{\partial A_{u_j}}{\partial r_i}B_{u_j}-\frac{\partial B_{u_j}}{\partial r_i}A_{u_j}\right)
\end{equation}
expressing mathematically its kinematics arises then from the association of Lie groups with the Poisson structures that  arise  on  the adjoint spaces  of the Lie algebras corresponding to the Lie groups \cite{arnold}. Using the same strategy in kinetic theory (here the Lie group is the group of canonical transformations $\mathbb{R}^6\rightarrow\mathbb{R}^6$), we arrive at the one particle distribution function as the state variable and the bracket (\ref{PB}) expressing its kinematics \cite{mar82}.

We  now discuss two examples of extensions: (i) extension to polymeric fluids and (ii) extension to fluid with inertia (the fluxes appearing on the right-hand side of (\ref{eqfm}) are seen as  extra state variables). We begin with the polymeric fluids. In the extension from simple to polymeric fluids, the microscopic particles from which the fluids are composed  acquire an internal structure. The simplest physical model of a macromolecule is a dumbbell \cite{bird2}. One point particle is extended to two-point particles connected by a spring. Leaving out  velocities, states of  macroscopic systems composed of  dumbbells are described by one-dumbbell distribution function $\psi(\rr,\RR)$, where $\RR$ is the vector representing the spring. Assuming that the velocities of the dumbbell particles change on  smaller scale than the scale of changes of  continuum, the kinematics of $\psi(\rr,\RR)$ (or alternatively its moments $c_{ij}(\rr)=\int d\rr R_iR_j\psi(\rr,\RR)$   is simply a passive advection (a Lie drag) of the dumbbell by the momentum field $\uu(\rr)$. The Poisson bracket is thus known. The spring inside the dumbbell determines the energy $E$. Knowledge of the bracket and the energy is enough to specify the constitutive relations for the non-dissipative part of the time evolution in isothermal polymeric fluids (see \cite{pkg} and references cited therein). 

The internal structure $\psi$ appears in the momentum flux $\JJ^{(u)}$ in the form of the  extra stress tensor and the momentum $\uu$ appears in  $\mathbb{R}$ as a force driving the internal structure.
Two independent calculations involving two independent collections of  assumptions about  mechanics of dumbbells immersed in continuum  are needed in the classical setting to obtain   the extra stress tensor and the force driving the internal structure \cite{bird2}.  Their  mutual compatibility is not guaranteed.

The second type of extension  is motivated by the form of the local conservation laws (\ref{eqfm}). The fields appearing on the right-hand side of (\ref{eqfm}) are candidates for the extra fields. From the mechanics point of view, this is a suggestion to extend Eqs.(\ref{eqfm}) by including inertia. Such extension
appears in fact already in the non-disipative Grad hierarchy \cite{grad,miroslav-grad}.  The first equation in the classical Grad hierarchy  is obtained  by integrating  the Boltzmann equation without the collision term over $\vv$, the second equation
by multiplying the Boltzmann equation without the collision term  by $\vv$ and integrating it over $\vv$, the third by multiplying it by $v_i$ and integrating over $\vv$, the fourth by multiplying it by $v_iv_j$ and integrating over $\vv$.  By continuing this procedure we obtain the non-dissipative Grad hierarchy governing the time evolution of infinite number of velocity moments of the one particle distribution function. The first five moments $(\int d\vv f(\rr,\vv), \int d\vv \vv f(\rr,\vv), \int d\vv \frac{\vv^2}{2m})f(\rr,\vv)$ are interpreted as the hydrodynamic fields $x^{(hyd)}$ and the moments $(\int d\vv \vv f(\rr,\vv), \int d\vv \vv\vv f(\rr,\vv), \int d\vv \vv\frac{\vv^2}{2m}f(\rr,\vv))$ as the fluxes $(\JJ^{(\rho)}, \JJ^{(u)}, \JJ^{(e)})$. Inclusion of the subsequent equations in the hierarchy is then interpreted as an extension of the classical hydrodynamic equations.

The first question that we ask is how can non-dissipative dynamics of an ideal gas can serve as a framework for non-dissipative mechanics of complex fluids. From the point of view of multiscale dynamics, the non-dissipative dynamics (the first term on the right-hand side of (\ref{generic})) of an ideal gas has two ingredients: Poisson bivector (expressing kinematics of one particle distribution function) and  energy of an ideal gas. The Poisson bivector can provide the framework but the very particular energy of an ideal gas certainly not. We can therefore try to construct the Grad hierarchy  with just the Poisson bivector and with an unspecified energy (we shall call it a Hamilton-Grad hierarchy). It turns out that such construction is  as easy as the construction of the classical Grad hierarchy.

It is  enough to restrict the Poisson bracket (\ref{PB}) to functions $A(f)$ and $B(f)$ that depend on $f$ only through the dependence on the velocity moments of $f$. We therefore replace $A_f$ in (\ref{PB}) with $A_{c^{(0)}}+v_iA_{c_i^{(1)}}+v_iv_jA_{c_{ij}^{(2)}}+...$, where $c_{i_1,i_2,...,i_k}^{(k)}(\rr)=\int d\vv v_{i_1}v_{i_2}...v_{i_k}f(\rr,\vv)$, and similarly we rewrite $B_f$. The resulting Grad hierarchy can be found in \cite{miroslav-grad}, \cite{pkg}.
We now make a few observations about the Hamilton-Grad  hierarchy.

First, we note that the Hamilton-Grad  hierarchy is different from the classical hierarchy. Second, we see that this new hierarchy is a Hamiltonian system of time evolution equations. Indeed, the Hamiltonian structure of the non-dissipative Boltzmann equation has been preserved. We have passed from the Poisson bracket (\ref{PB}) to the Poisson bracket for the moments only by restricting the functions $A(f)$ and $B(f)$ to a particular class of functions. 

The Hamilton-Grad hierarchy is still an infinite hierarchy. Can we close it? By a closure we mean a transformation of the infinite hierarchy into a finite hierarchy of equations governing the time evolution of a finite number moments $c^{(0)}(\rr),...,c^{(n)}(\rr)$  and  remaining infinite hierarchy governing the time evolution of the remaining infinite number of moments $c^{(n+1)}(\rr), c^{(n+2)}(\rr),....$. More specifically, we ask: can we close the infinite Hamilton-Grad hierarchy while keeping its Hamiltonian structure? In particular, we are interested in finding a hierarchy of  Hamiltonian time evolution equations governing the time evolution of $c^{(0)}(\rr),...,c^{(n)}(\rr)$. The closure is obviously related to the presence of dissipation. We can however still make a few observations about the Hamilton-Grad hierarchy and about the closure while considering only the non-dissipative part of the time evolution.

The fluxes that arise on the right-hand side of the time evolution equations in the classical Grad hierarchy are the same as the fluxes on their left-hand side. In other words, the extra fields adopted as extra state variables are the fluxes arising in the vector fields. This is not the case in the Hamilton-Grad hierarchy. The fluxes in the vector fields of the Hamilton-Grad hierarchy are conjugates $c^{*(j)}=E_{c^{(j)}}(c^{(0)},c^{(1)},...c^{(n)})$. The identification of  fluxes with extra fields is a consequence of the ideal-gas limitation of the classical Grad hierarchy.

Since we  are restricting ourselves to  non-dissipative time evolution, the second law (\ref{2law}) becomes $\frac{\partial s(\rr)}{\partial t}=-\partial_iJ^{(s)}_i(\rr)$. We have already seen
in the context of the Boltzmann equation  that $\frac{\partial S}{\partial t}=0$ for $S=\int d\rr \int d\vv \eta(f(\rr,\vv))$, where $\eta:\mathbb{R}\rightarrow \mathbb{R}$. We choose the entropy fields $s(\rr)=\int d\vv\eta(f(\rr,\vv))$. Straightforward  calculations (see \cite{miroslav-grad},\cite{pkg}) lead to
\begin{eqnarray}
\frac{\partial s(\rr)}{\partial t} &=& -\frac{\partial J^{(s)}_{\alpha}(\rr)}{\partial r_{\alpha}}\nonumber \\
J^{(s)}_{\alpha}&=& b^{(0)}c^{*(1)}_{\alpha}+2b^{(1)}_ic^{*(2)}c^{(2)}_{\alpha i} + 3b^{(2)}_{ij}c^{*(3)}_{\alpha ij}+.... \nonumber \\
b^{(0)}(\rr)&=&\int d\vv \eta(f(\rr,\vv))=s(\rr)\nonumber \\
b^{(k)}_{i_1,...,i_k}(\rr)&=&\int d\vv v_{i_1}...v_{i_k}\eta(f(\rr,\vv)).
\end{eqnarray}
The second law holds for the Hamilton-Grad hierarchy and the entropy flux is explicitly expressed as a function of the entropy moments $\bb(\rr)$ and the conjugate state variables $\cc^*$.

Now we turn to  the closure that keeps the Hamiltonian structure. It is easy to verify (see \cite{momentum-Euler}, \cite{pkg}) that by choosing $\hat{x}^{(hyd)}=(c^{(0)}(\rr), c^{(1)}(\rr),s(\rr))$ as state variables, the Poisson bracket generating the Hamilton-Grad hierarchy splits into two Poisson brackets, one expressing kinematics of  $\hat{x}^{(hyd)}$ and the other expressing kinematics of the remaining moments $\cc$ (the entropy field $s(\rr)=\int d\vv\eta(f(\rr,\vv))$). Moreover, the hydrodynamic fields form a subalgebra of the infinite Grad Lie algebra \cite{momentum-Euler}.

The former Poisson bracket is
\begin{eqnarray}\label{PBs}
\{A,B\}^{(hyd)}&=&\int d\rr \rho\left(\frac{\partial A_\rho}{\partial r_i}B_{u_i}-\frac{\partial B_\rho}{\partial r_i}A_{u_i}\right)\nonumber \\
&&+\int d\rr u_i\left(\frac{\partial A_{u_i}}{\partial r_j}B_{u_j}-\frac{\partial B_{u_i}}{\partial r_j}A_{u_j}\right)\nonumber \\
&&+\int d\rr s\left(\frac{\partial A_{s}}{\partial  r_i}B_{u_i}-\frac{\partial B_{s}}{\partial r_i}A_{u_i}\right) 
\end{eqnarray}
and generating by it the non-dissipative equations of the classical non-equilibrium thermodynamics
\begin{eqnarray}\label{fms}
\frac{\partial \rho}{\partial t}&=& -\partial_i(\rho E_{u_i})\nonumber \\
\frac{\partial u_i}{\partial t}&=& -\partial_j(u_i E_{u_j})-\rho \partial_i E_\rho -s\partial_i E_s\nonumber \\
\frac{\partial s}{\partial t}&=& -\partial_i(s E_{u_i}).
\end{eqnarray}
In this way we are proving that the non-dissipative part of the classical non-equilibrium thermodynamics is Hamiltonian and is the same as the non-dissipative part of the multiscale non-equilibrium thermodynamics with the state variables $\hat{x}^{(hyd)}=(c^{(0)}(\rr), c^{(1)}(\rr),s(\rr))$.

Is there any other way to split the Poisson bracket expressing kinematics of the infinite number of moments  into two Poisson brackets, one  expressing kinematics of a finite number of moments $c^{(0)}(\rr),...,c^{(n)}(\rr)$ and the second kinematics of $c^{(n+1)}(\rr), c^{(n+2)}(\rr),....$?  Two arguments supporting the negative answer have appeared in \cite{momentum-Euler}. The first one arises from  an analysis of the Lie algebras associated with the brackets and the second from arguing that the existence of an extended hydrodynamics with the state variables $c^{(0)}(\rr),...,c^{(n)}(\rr)$ would contradict  experimental observations of turbulence.

\subsection{Local equilibrium assumption}

Multiscale view of non-equilibrium thermodynamics shows how are the second law, the local equilibrium assumption, and the Hamiltonian structure of the non-dissipative part of the time evolution intertwined. Their interdependence  begins already with choosing $\hat{x}^{(hyd)}=(c^{(0)}(\rr), c^{(1)}(\rr),s(\rr))$ instead of
$x^{(hyd)}=(c^{(0)}(\rr), c^{(1)}(\rr),\epsilon(\rr))$. The former choice guarantees the second law and that fact that the latter choice is the choice coming from the classical mechanics-based considerations it is tacitly assumed that there is a one-to-one relation between $s(\rr)$ and $\epsilon(\rr)$. This is indeed the case if the relation $\frac{\partial \epsilon}{\partial s}=T>0$, where $T$ is the absolute temperature,  holds. But this relation is one of the relations implied by the local equilibrium.
This means that already the choice of $\hat{x}^{(hyd)}$ as state variables relates  (recall that this choice is dictated in the multiscale nonequilibrium thermodynamics by the closure of the infinite hierarchy) relates  the second law, the local equilibrium assumption, and the Hamiltonian structure of the nondissipative part of the time evolution. 

Still another such relation is the relation between the pressure $p(\rr)$ and $x^{(hyd)}$ appearing in (\ref{PBs}). We conclude this comment by noting that there is a fundamental difference between extensions with some extra fields $\xi^{(exthyd)}(\rr)$  that involve the internal energy field $\epsilon(\rr)$ or the entropy field $s(\rr)$ as one of the state variables and the extension in which both $\epsilon(\rr)$ and $s(\rr)$ are functions of the state variables determined by constitutive relations. The former extension includes tacitly the local equilibrium assumption (one-to-one relation between the fields $\epsilon(\rr)$ and $s(\rr)$ is tacitly assumed).

The second law, the local equilibrium assumption, and the Hamiltonian structure are also closely related to the mathematical regularity of solutions of the system of local conservation  laws of the type (\ref{eqfm}).

Another feature of Hamiltonian continuum mechanics is the automatic conservation of energy, which follows from the skew-symmetry of the underlying Poisson bracket. In numerical mathematics, this features if often called the compatibility of the energy flux \cite{god} and is needed for proofs of symmetric hyperbolicity of conservation laws \cite{shtc-generic}.

\subsection{Nonlinear dissipation,  Braun-Le Ch{\^a}telier principle}
The dissipative part of the GENERIC framework is generated by a dissipation potential and entropy. The dissipation potential is quadratic near the thermodynamic equilibrium, but far from equilibrium it can be non-linear and even non-convex \cite{nonconvex}. A statistical derivation of gradient dynamics is based on the principle of large deviations \cite{mielke-potential}, where gradient dynamics represents the most probable trajectory of a stochastic process. 

Moreover, gradient dynamics can be seen as a generalization of the Onsager reciprocal relations, as derivatives of the dissipation potential are exchangable. The Hamiltonian part then provides the skew-symmetric coupling with respect to the simultaneous time-reversal transformation and transposition \cite{pkg}. 

The convexity of the dissipation potential is related to the stability of the thermodynamic equilibrium. The Braun-Le Ch\^{a}telier principle \cite{lechatelier,landau5}, which states that a perturbation of a stable state leads to changes that counteract the perturbation, can be generalized far from equilibrium \cite{BLC-dissipative}. 

For instance, when we have two chemical reactions, one fast $(A->B)$, one slow $(B->C)$, both prescribed by gradient dynamics, so that the total dissipation potential is the sum of the two particular dissipation potentials, 
\begin{subequations}
\begin{equation}
\Xi^{tot} = \Xi^{fast} + \Xi^{slow}
\end{equation}
where 
\begin{equation}
  \Xi^{fast} = W^fast(c_A, c_B) \left( \cosh\left(\frac{X_{fast}}{2k_B}\right) - 1 \right), \quad X_{fast} = c^*_B - c^*_A
\end{equation}
and
\begin{equation}
  \Xi^{slow} = W^{slow}(c_B, c_C) \left( \cosh\left(\frac{X_{slow}}{2k_B}\right) - 1 \right), \quad X_{slow} = c^*_C - c^*_B,
\end{equation}
see \cite{grchemkin,pkg}.
\end{subequations}

The evolution equations are then 
\begin{subequations}\label{eq.chem.gen}
\begin{align}
  \dot{c}_A &= \frac{\partial \Xi^{tot}}{\partial c^*_A} = \frac{\partial \Xi^{fast}}{\partial c^*_A}
  = \frac{W^{fast}}{2k_B}\sinh(X_{fast}/{2k_B}) \\
  \dot{c}_B &= \frac{\partial \Xi^{tot}}{\partial c^*_B} = \frac{\partial \Xi^{fast}}{\partial c^*_B} + \frac{\partial \Xi^{slow}}{\partial c^*_B}\nonumber\\
  &= -\frac{W^{fast}}{2k_B}\sinh(X_{fast}/{2k_B}) + \frac{W^{slow}}{2k_B}\sinh(X_{slow}/{2k_B}) \\
  \dot{c}_C &= \frac{\partial \Xi^{tot}}{\partial c^*_C} = \frac{\partial \Xi^{slow}}{\partial c^*_C}
  = -\frac{W^{slow}}{2k_B}\sinh(X_{slow}/{2k_B}).
\end{align}
\end{subequations}
The conjugate concentrations are given by the corresponding derivatives of the entropy, 
\begin{equation}
  S = -k_B \int d\rr \left( c_A(\ln(c_A/\bar{c}) - 1) + c_B(\ln(c_B/\bar{c}) - 1) + c_C(\ln(c_C/\bar{c}) - 1) \right),
\end{equation}
where $\bar{c}$ is a reference concentration, so
\begin{equation}
  c^*_A = -k_B \ln\left(\frac{c_A}{\bar{c}}\right), \quad c^*_B = -k_B \ln\left(\frac{c_B}{\bar{c}}\right), \quad c^*_C = -k_B \ln\left(\frac{c_C}{\bar{c}}\right).
\end{equation}
When the reaction kernels $W^{fast}$ and $W^{slow}$ are chosen as 
\begin{equation}
  W^{fast} = 4k_B k^{fast}\sqrt{ c_A c_B} , \quad W^{slow} = 4k_B k^{slow}\sqrt{ c_B c_C},
\end{equation}
the evolution equations become
\begin{subequations}
\begin{align}
  \dot{c}_A &= -k^{fast}(c_A - c_B) \\
  \dot{c}_B &= k^{fast}(c_A - c_B) - k^{slow}(c_B - c_C) \\
  \dot{c}_C &= k^{slow}(c_B - c_C), 
\end{align}
that is the classical mass action law. 
\end{subequations}

The dissipative Braun-Le Ch\^{a}telier principle \cite{BLC-dissipative} then tells that when the slow force $c^*_C$ is perturbed, it leads to a perturbation of the fast force $c^*_B$. After the $c_B$ has relaxed, the dissipation potential $\Xi^{slow}$ has lower second derivative, and is thus less steep, and the rate $\dot{c}_C$ is smaller than it was just after the perturbation. Mathematically, this is expressed as
\begin{equation}
  \left(\frac{\partial \dot{c}_C}{\partial c^*_C}\right)_{c^*_B \text{ fixed}} > \left(\frac{\partial \dot{c}_C}{\partial c^*_C}\right)_{\dot{c}_B = 0}.
\end{equation}
When we approximate the $\sinh$ functions in Equations \eqref{eq.chem.gen} by their first-order Taylor expansions, we obtain linear gradient dynamics,
\begin{subequations}
\begin{align}
  \dot{c}_A &= \frac{W^{fast}}{4k_B^2} (c^*_B - c^*_A) \\
  \dot{c}_B &= \frac{W^{fast}}{4k_B^2} (c^*_A - c^*_B) + \frac{W^{slow}}{4k_B^2} (c^*_C - c^*_B) \\
  \dot{c}_C &= \frac{W^{slow}}{4k_B^2} (c^*_B - c^*_C).
\end{align}
\end{subequations}
From the first two equations, we obtain that
\begin{equation}
  \dot{c}_B = \frac{W^{fast}}{4k_B^2} c^*_A  + \frac{W^{slow}}{4k_B^2} c^*_C - \left(\frac{W^{fast}}{4k_B^2} + \frac{W^{slow}}{4k_B^2}\right) c^*_B,
\end{equation}
or
\begin{equation}
c^*_B = \frac{\frac{W^{fast}}{4k_B^2} c^*_A  + \frac{W^{slow}}{4k_B^2} c^*_C - \dot{c}_B}{\frac{W^{fast}}{4k_B^2} + \frac{W^{slow}}{4k_B^2}}
\end{equation}
Then, the equation for $c_C$ can be reformulated with the fast rates $\dot{c}_B$ and $\dot{c}_A$ as
\begin{equation}
  \dot{c}_C = \frac{\frac{W^{fast}}{4k_B^2} c^*_A  + \frac{W^{slow}}{4k_B^2} c^*_C - \dot{c}_B}{\frac{W^{fast}}{W^{slow}} + 1} - \frac{W^{slow}}{4k_B^2} c^*_C,
\end{equation}
and the derivative of $\dot{c}_C$ with respect to $c^*_C$ at constant $\dot{c}_B$ is
\begin{equation}
  \left|\left(\frac{\partial \dot{c}_C}{\partial c^*_C}\right)_{\dot{c}_B}\right| = 
  \left|\frac{W^{slow}}{4k_B^2}\left(\frac{1}{1+\frac{W^{fast}}{W^{slow}}}-1\right)\right|
  < \frac{W^{slow}}{4k_B^2} = \left|\left(\frac{\partial \dot{c}_C}{\partial c^*_C}\right)_{c^*_B}\right|. 
\end{equation}
In other words, the change of the rate $\dot{c}_C$ with respect to the force $c^*_C$ is smaller after the fast variable $c_B$ has relaxed than it would be if $c_B$ was kept fixed. This is an illustration of the Braun-Le Ch\^{a}telier principle in dissipative thermodynamics.

\section{Conclusion}
We have presented a brief review of the multiscale non-equilibrium thermodynamics and discussed a some contributions that the multiscale viewpoint brings to the classical and extended non-equilibrium thermodynamics. The multiscale viewpoint is based on the recognition that the time evolution of macroscopic systems is often governed by two independent parts, one expressing kinematics and the other thermodynamics. The kinematics is expressed in terms of Poisson brackets and the thermodynamics in terms of dissipation potentials and entropy. The multiscale viewpoint is abstract and can be applied to many different mesoscopic levels of description. The classical and extended non-equilibrium thermodynamics are just particular realizations of the multiscale non-equilibrium thermodynamics. 

Eventually, the multiscale framework brings also geometry, hyperbolicity, and generalized stability conditions, for instance the dissipative Braun-Le Ch\^{a}telier principle. The latter is illustrated here on a simple example of two chemical reactions, but it can be applied to more complex systems, for instance to viscoelastic fluids \cite{BLC-dissipative}.

\section*{Acknowledgment}
MP was supported by Czech Science Foundation, project 23-05736S. 
MP is a member of the Nečas Center for Mathematical Modeling. 

\newcommand{\etalchar}[1]{$^{#1}$}


\begin{thebibliography}{EGGP19}

\bibitem[Arn66]{arnold}
V.I. Arnold.
\newblock Sur la g\'{e}ometrie diff\'{e}rentielle des groupes de lie de
  dimension infini et ses applications dans l'hydrodynamique des fluides
  parfaits.
\newblock {\em Annales de l'institut Fourier}, 16(1):319--361, 1966.

\bibitem[BHAC87]{bird2}
R.~B. Bird, O.~Hassager, R.~C. Armstrong, and C.~F. Curtiss.
\newblock {\em Dynamics of Polymeric Fluids}, volume~2.
\newblock Wiley, New York, 1987.

\bibitem[Cal60]{callen}
H.B. Callen.
\newblock {\em Thermodynamics: an introduction to the physical theories of
  equilibrium thermostatics and irreversible thermodynamics}.
\newblock Wiley, 1960.

\bibitem[Cha84]{lechatelier}
Henry~Le Chatelier.
\newblock Sur un \'{e}nonc\'{e} g\'{e}n\'{e}ral des lois des \'{e}quilibres
  chimiques.
\newblock {\em Comptes-rendus de l'Acad\'{e}mie des sciences}, 99:786--789,
  1884.

\bibitem[Cle95]{clebsch}
A.~Clebsch.
\newblock \"{U}ber die {I}ntegration der {H}ydrodynamische {G}leichungen.
\newblock {\em Journal f\"{u}r die reine und angewandte Mathematik}, 56:1--10,
  1895.

\bibitem[dGM84]{dgm}
S.~R. de~Groot and P.~Mazur.
\newblock {\em Non-equilibrium Thermodynamics}.
\newblock Dover Publications, New York, 1984.

\bibitem[DV80]{dv}
I.~E. Dzyaloshinskii and G.~E. Volovick.
\newblock Poisson brackets in condense matter physics.
\newblock {\em Annals of Physics}, 125(1):67--97, 1980.

\bibitem[EGGP19]{momentum-Euler}
O{\u g}ul Esen, Miroslav Grmela, Hasan G{\u u}mral, and Michal Pavelka.
\newblock Lifts of symmetric tensors: {F}luids, plasma, and grad hierarchy.
\newblock {\em Entropy}, 21(9):907, 2019.

\bibitem[Eul55]{euler}
L.~Euler.
\newblock Principes g\'{e}n\'{e}raux du mouvement des fluides.
\newblock {\em Acad\'{e}mie Royale des Sciences et des Belles-Lettres de
  Berlin, M\'{e}moires}, 11, 1755.
\newblock English translation is available in Physica D 237, 1825-1839 (2008).

\bibitem[Ges83]{boltzmann}
L.B. Gesamtausgabe.
\newblock {\em Ludwig Boltzmann Gesamtausgabe - Collected Works}.
\newblock 1983.

\bibitem[GHJ{\etalchar{+}}17]{miroslav-grad}
Miroslav Grmela, Liu Hong, David Jou, Georgy Lebon, and Michal Pavelka.
\newblock Hamiltonian and godunov structures of the grad hierarchy.
\newblock {\em Physical Review E}, 95(033121), 2017.

\bibitem[GO97]{go}
Miroslav Grmela and Hans~Christian \"{O}ttinger.
\newblock Dynamics and thermodynamics of complex fluids. {I}. {D}evelopment of
  a general formalism.
\newblock {\em Phys. Rev. E}, 56:6620--6632, Dec 1997.

\bibitem[God72]{god}
S.K. Godunov.
\newblock Symmetric form of the magnetohydrodynamic equation.
\newblock {\em Chislennye Metody Mekhaniki Sploshnoi Sredy}, 3(1):26--34, 1972.

\bibitem[Gra58]{grad}
H.~Grad.
\newblock {\em Encyclopedia of Physics}, volume~12, chapter Principles of
  Kinetic Theory of Gases.
\newblock Springer-Verlag, Berlin, 1958.

\bibitem[Grm84]{grcontmath}
M.~Grmela.
\newblock Particle and bracket formulations of kinetic equations.
\newblock {\em Contemporary Mathematics}, 28:125--132, 1984.

\bibitem[Grm12]{grchemkin}
M.~Grmela.
\newblock Fluctuations in extended mass-action-law dynamics.
\newblock {\em Physica D Nonlinear Phenomena}, 241:976--986, May 2012.

\bibitem[JCVL10]{jou-eit}
D.~Jou, J.~Casas-Vázquez, and G.~Lebon.
\newblock {\em Extended Irreversible Thermodynamics}.
\newblock Springer-Verlag, New York, 4th edition, 2010.

\bibitem[J.E82]{mar82}
Marsden J.E.
\newblock A group theoretical approach to the equations of plasma physics.
\newblock {\em Canad. Math. Bull.}, 25:129--142, 1982.

\bibitem[JP18]{nonconvex}
Adam Jane{\v c}ka and Michal Pavelka.
\newblock Non-convex dissipation potentials in multiscale non-equilibrium
  thermodynamics.
\newblock {\em Continuum Mechanics and Thermodynamics}, 30(4):917--941, 2018.

\bibitem[JR11]{david-liliana}
D.~Jou and L.~Restuccia.
\newblock Mesoscopic transport equations and contemporary thermodynamics: an
  introduction.
\newblock {\em Contemporary Physics}, 52(5):465--474, 2011.

\bibitem[Kau84]{kauf}
A.N. Kaufman.
\newblock Dissipative hamiltonian systems: A unifying principle.
\newblock {\em Physics Letters A}, 100(8):419--422, 1984.

\bibitem[Liu72]{liu1972}
I-Shih Liu.
\newblock Method of lagrange multipliers for exploitation of the entropy
  principle.
\newblock {\em Archive for Rational Mechanics and Analysis}, 46(2):131--148,
  1972.

\bibitem[LL69]{landau5}
L.D. Landau and E.M. Lifschitz.
\newblock {\em Statistical physics}.
\newblock Number pt. 1 in Course of theoretical physics. Pergamon Press, 1969.

\bibitem[MG80]{morrison-greene1980}
Philip~J. Morrison and John~M. Greene.
\newblock Noncanonical hamiltonian density formulation of hydrodynamics and
  ideal magnetohydrodynamics.
\newblock {\em Phys. Rev. Lett.}, 45:790--794, Sep 1980.

\bibitem[Mor84]{mor}
P.~J. Morrison.
\newblock Bracket formulation for irreversible classical fields.
\newblock {\em Physics Letters A}, 100(8):423--427, 1984.

\bibitem[MPR14]{mielke-potential}
A.~Mielke, M.~A. Peletier, and D.~R.~M. Renger.
\newblock On the relation between gradient flows and the large-deviation
  principle, with applications to {M}arkov chains and diffusion.
\newblock {\em Potential Analysis}, 41(4):1293--1327, 2014.

\bibitem[OG97]{og}
Hans~Christian \"Ottinger and Miroslav Grmela.
\newblock Dynamics and thermodynamics of complex fluids. {II}. {I}llustrations
  of a general formalism.
\newblock {\em Phys. Rev. E}, 56:6633--6655, Dec 1997.

\bibitem[{\"O}tt05]{hco}
H.C. {\"O}ttinger.
\newblock {\em Beyond Equilibrium Thermodynamics}.
\newblock Wiley, New York, 2005.

\bibitem[PG19]{BLC-dissipative}
M.~Pavelka and M.~Grmela.
\newblock Braun-{L}e {C}hatelier principle in dissipative thermodynamics.
\newblock {\em Proceedings of Accademia Peloritana dei Pericolanti},
  97(S1):A22, 2019.

\bibitem[PKG18]{pkg}
Michal Pavelka, V{\' a}clav Klika, and Miroslav Grmela.
\newblock {\em Multiscale Thermo-Dynamics}.
\newblock de Gruyter, Berlin, 2018.

\bibitem[Poi09]{poisson}
S.~Poisson.
\newblock Sur la variation des constantes arbitraires dans les questions de
  mécanique.
\newblock {\em J. Ecole Polytechn.}, 8:266--344, 1809.

\bibitem[PPRG18]{shtc-generic}
Ilya Peshkov, Michal Pavelka, Evgeniy Romenski, and Miroslav Grmela.
\newblock Continuum mechanics and thermodynamics in the {H}amilton and the
  {G}odunov-type formulations.
\newblock {\em Continuum Mechanics and Thermodynamics}, 30(6):1343--1378, 2018.

\bibitem[Pri55]{prigogine-tip}
I.~Prigogine.
\newblock {\em Introduction to Thermodynamics of Irreversible Processes}.
\newblock Thomas, New York, 1955.

\bibitem[Wal67]{waldmann}
L.~Waldmann.
\newblock Non-equilibrium thermodynamics of boundary conditions.
\newblock {\em Zeitschrift f{\" u}r Naturforschung A}, 22(8):1269--1280, 1967.

\end{thebibliography}
\end{document}